\documentclass[12pt]{iopart}

\usepackage{epstopdf}
\usepackage{graphicx}
\usepackage{color}
\usepackage{bm}
\usepackage{setspace}
\usepackage{pgfplots}
\usepackage{comment}
\usepackage{subfigure}
\usepackage{amsfonts}
\usepackage{dcolumn}

\usepackage[colorlinks=true,linkcolor=blue,citecolor=blue]{hyperref}
\hypersetup{colorlinks=true,citecolor=blue,linkcolor=red,urlcolor=red}

\begin{document}

\renewcommand{\thefigure}{\arabic{figure}}
\def\be{\begin{equation}}
\def\ee{\end{equation}}
\def\ber{\begin{eqnarray}}
\def\eer{\end{eqnarray}}

\def\kv{{\bf k}}
\def\qv{{\bf q}}
\def\pv{{\bf p}}
\def\sigmav{{\bf \sigma}}
\def\tauv{{\bf \tau}}
\newcommand{\h}[1]{{\hat {#1}}}
\newcommand{\hdg}[1]{{\hat {#1}^\dagger}}
\newcommand{\bra}[1]{\left\langle{#1}\right|}
\newcommand{\ket}[1]{\left|{#1}\right\rangle}
\newcommand{\braket}[2]{\ensuremath{\left\langle{#1}\middle|{#2}\right\rangle}}

\title{Adiabatic state preparation of stripe phases with strongly magnetic
atoms}

\date{\today}

\author{Azadeh Mazloom$^{1,2,3}$, Beno\^it Vermersch$^{2, 3}$, Mikhail A. Baranov$^{2, 3}$ and Marcello Dalmonte$^{2, 4}$}

\address{$^1$Department of Physics, Institute for Advanced Studies in Basic Sciences
(IASBS), Zanjan 45137-66731, Iran}

\address{$^2$Institute for Theoretical Physics, University of Innsbruck, A-6020
Innsbruck, Austria}

\address{$^3$Institute for Quantum Optics and Quantum Information of the Austrian
Academy of Sciences, A-6020 Innsbruck, Austria}

\address{$^4$Abdus Salam International Centre for Theoretical Physics, Strada
Costiera 11, Trieste, Italy}
\begin{abstract}
We propose a protocol for realising the stripe phase in two spin models
on a two-dimensional square lattice, which can be implemented with
strongly magnetic atoms (Cr, Dy, Er, etc.) in optical lattices by
encoding spin states into Zeeman sublevels of the ground state manifold.
The protocol is tested with cluster-mean-field time-dependent variational
ans\"atze, validated by comparison with exact results for small systems,
which enable us to simulate the dynamics of systems with up to 64
sites during the state-preparation protocol. This allows, in particular,
to estimate the time required for preparation of the stripe phase
with high fidelity under real experimental conditions. 
\end{abstract}

\section{Introduction}

Experiments with ultra-cold quantum gases of atoms and molecules have
recently witnessed tremendous progresses in realising and probing
many-body physics with long-range interacting spin systems~\cite{Baranov}.
Several works have reported the observation of coherent dynamics with
both magnetic and Rydberg atoms~\cite{Lahaye2009,Browaeys} and polar molecules~\cite{Baranov},
ranging from the time evolution of spin models after a quantum quench~\cite{PazPRL,Yan,Malossi2014,Schauss2015},
to the realisation of Hubbard and extended Bose-Hubbard models featuring both
local and non-local interactions~\cite{Parsons,Boll,Cheuk,Baier}. These experimental
studies are paving the way to the investigation of quantum magnetism
in atomic and molecular quantum systems, in particular, about the
stability of ordered magnetic phases in frustrated systems~\cite{Baranov,Lahaye2009}.

One example of such phases is the stripe phase - supporting ferromagnetic
order along one direction, and antiferromagnetic along the perpendicular
one. Stripe order has been widely discussed in quantum magnetism,
both in the context of spin (Heisenberg and Ising type)~\cite{Lacroix,Jin}
and fermionic (Hubbard and t-J) models (with possible implications
on the finite-doping phases of high-temperature superconductors)~\cite{Lee2006}.
Moreover, stripe ordering represents a natural counterpart to anti-ferromagnets
in the presence of angular dependent interactions~\cite{Lacroix}:
these interactions do naturally occur in dipolar gases of magnetic
atoms such as Cr, Dy, and Er, all of which are by now being observed
in quantum degenerate regimes~\cite{Pasquiou1,Frisch2,Ferrier2016,Lahaye2007,Lu2012}.
The main virtues of such atomic species is that they can combine dipolar
interactions with close-to-ideal initial state preparation (such as
Mott insulators with unit filling) and long coherence times (of the order
of 1 $s$) which compare well with the interaction timescales (typically
of the order of 10 to 100 Hz in units of $\hbar$)~\cite{Baranov,Lahaye2009}.
These two features make magnetic atoms an ideal platform for the realisation
and investigation of stripe phases in atomic many-body systems, an
approach very much complementary to recent experimental findings~\cite{Li2016}
in the context of spin orbit coupled Bose gases~\cite{Zhang2012,Li2012}.

One of the key questions toward the realisation of such magnetic phases
is the identification of clear-cut state preparation protocols which
can warrant experimental observability of specific signatures (i.e.
correlation functions) under realistic experimental conditions. In
this work, we address the preparation of the stripe phase in spin-1
Heisenberg-type and spin-$1/2$ Ising-type models on the square lattice,
with antiferromagnetic interactions between neighbours in one direction,
and ferromagnetic along the other, which can be realised with magnetic
atoms trapped in optical lattices. For both models, the proposed adiabatic
protocols for preparing the stripe phase are tested using a time-dependent
variational principle, directly inspired by (cluster) mean field theory.
This method agrees well with the exact treatment of small systems,
and allows us to tackle the full-time dependent dynamics for lattices
containing of the order of 100 sites - which is comparable to the
largest, defect-free Mott insulator states available in current experiments~\cite{Gross2014}.
Our results strongly indicate that the realisation of the stripe phases
is within the reach of current experiments with strongly magnetic
atoms such as Dy or Er, with strong signatures of the stripe formation
already visible for moderate speeds of changes in the system's parameters
during the preparation protocols. Beyond the magnetic atom implementation
discussed here, our findings are also applicable to other physical
systems, such as polar molecules and Rydberg atoms in optical lattices~\cite{Lahaye2009},
and arrays of superconducting circuits~\cite{Dalmonte2015,Viehmann},
which are described by very similar lattice spin Hamiltonians.

The structure of the paper goes as follows. In section~\ref{sec:theory},
we discuss the two dipolar spin models of interest. In section~\ref{sec:result},
we introduce the time-dependent variational principle, and discuss
in detail the state preparation protocols for stripe phases in both
spin-$1/2$ and spin-1 models. Section~\ref{sec:imple} contains
two implementation schemes using magnetic atoms in optical lattices,
and provides a glimpse of the corresponding experimental timescales
for both coherent and incoherent (dissipation) dynamics. Finally,
we draw our conclusions in section~\ref{sec:conclusion}.


\section{Model Hamiltonians}

\label{sec:theory} In this section we introduce two spin models of
interacting dipoles whose phase diagrams include stripe phases. The
setup we have in mind is shown in figure~\ref{fig:lattice}(a). It
contains magnetic atoms at fixed positions on a square lattice, interacting
via dipole-dipole interactions (DDI) and with an external magnetic
field along the $z$-axis. This is the common starting point for the realisation of both spin-$1/2$
and spin-1 lattice models. In the following we discuss the corresponding
Hamiltonians and the phase diagrams. For the implementation of the
models, we refer the reader to section \ref{sec:imple}.

\subsection{Spin-$1/2$ model}

\label{sec:spin_half}

\begin{figure}
\centering \includegraphics[width=0.6\textwidth]{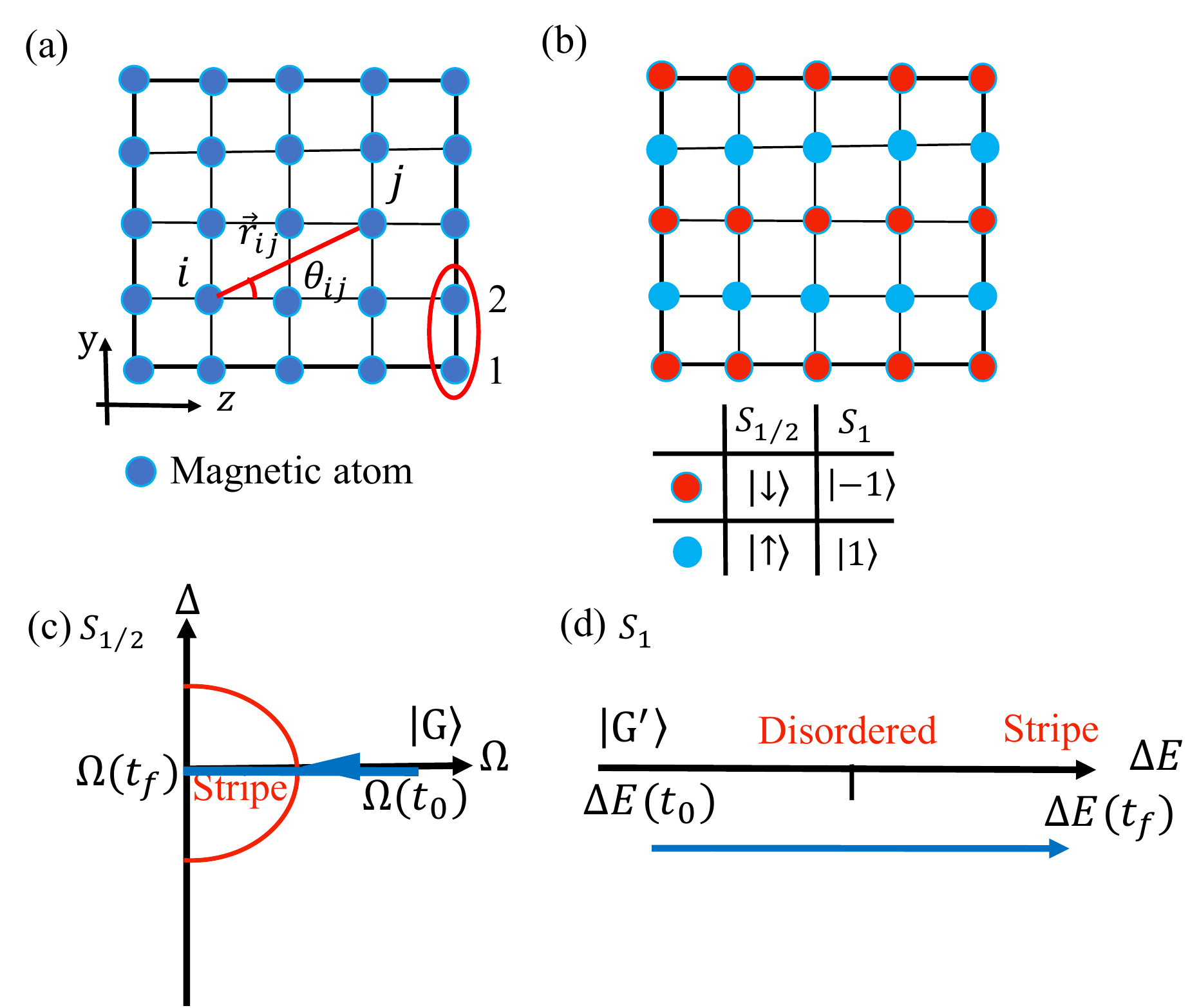}
\caption{(a) The arrangement of dipoles on a square lattice with a single atom
per site, $\vec{r}_{ij}$ is the relative vector between spins $i$
and $j$, $\theta_{ij}$ is the angle between $\vec{r}_{ij}$ and
the quantisation axis z. (b) Representation of the stripe phase formed
by aligned dipoles in direction z and anti-aligned ones in direction
y. (c) The schematic phase diagram of the ground state of the spin-$1/2$
model in $\Delta-\Omega$ space. The blue line shows the sweep path
along which the ground state of the system adiabatically changes from
a ferromagnetic phase $\protect\ket G$ at $\Omega(t_{0})\gg|C_{ij}|$
and $\Delta(t_{0})=0$ to the stripe phase at $\Omega(t_{f})=0$ and
$\Delta(t_{f})=0$. (d) Different phases of the spin-1 model as a
function of $\Delta E$. The stripe phase can be prepared by a slow
change of $\Delta E$ along the sweep path shown as the blue arrow.
\label{fig:lattice}}
\end{figure}

The first model we consider is obtained by assuming the two involved
atomic states as states
$\ket{\uparrow},\ket{\downarrow}$ of pseudo-spin $1/2$ particles.
The corresponding Hamiltonian reads 
\begin{equation}
H=\sum_{i<j}^{N}C_{ij}\sigma_{z}^{(i)}\sigma_{z}^{(j)}+\Omega\sum_{i}^{N}\sigma_{x}^{(i)}-\Delta\sum_{i}^{N}\sigma_{z}^{(i)},
\label{eq:tot_hamiltonian_2level}
\end{equation}
where $N$ is the number of atoms, $\sigma_{z}^{(i)}=\ket{\uparrow}_{i}\bra{\uparrow}_{i}-\ket{\downarrow}_{i}\bra{\downarrow}_{i}$
and $\sigma_{x}^{(i)}=\ket{\uparrow}_{i}\bra{\downarrow}_{i}+\ket{\downarrow}_{i}\bra{\uparrow}_{i}$
are the Pauli matrices acting at site $i$. The first term in this
Hamiltonian describes the Ising-type DDI between two spins on sites
$i$ and $j$ with $C_{ij}=c_{d}(1-3\cos^{2}\theta_{ij})/r_{ij}^{3}$,
where the interaction constant $c_{d}$ depends on the specific implementation
of the spin models, $r_{ij}=|\textbf{r}_{i}-\textbf{r}_{j}|$ is the
distance (in units of the lattice spacing $a$) between the sites,
and $\theta_{ij}$ the corresponding angle with respect to the quantisation
axis $z$ {[}c.f. figure~\ref{fig:lattice}(a){]}. In order to have
the stripe phase as the ground state, we place the quantisation axis in the plane of the atoms such that DDI is attractive in one
direction ($\theta_{ij}=0$) and repulsive in the other direction
($\theta_{ij}=\pi/2$)~\footnote{If the quantisation axis is perpendicular to the lattice, the angle
$\theta_{ij}$ is fixed to $\pi/2$ for all atomic pairs.}. Finally, the last two terms in equation~(\ref{eq:tot_hamiltonian_2level})
correspond to transverse ($\Omega$) and longitudinal ($\Delta$)
magnetic fields, which will be used for the state-preparation.

The zero-temperature phase diagram of the Hamiltonian (\ref{eq:tot_hamiltonian_2level})
is schematically shown in figure~\ref{fig:lattice}(c). It hosts
a stripe phase for $\left|\Omega\right|,\,\left|\Delta\right|$$<\left|c_{d}\right|$
with spins aligned along the $z$-axis and anti-aligned along the
$y$-axis {[}c.f. figure~\ref{fig:lattice}(b){]}. This phase is
destroyed by increasing magnetic fields when $\left|\Omega\right|,\,\left|\Delta\right|$$\sim\left|c_{d}\right|$
(see~\cite{Weimer,Sela,Vermersch} in the context of Rydberg atoms),
and for $\left|\Omega\right|,\,\left|\Delta\right|$$\gg\left|c_{d}\right|$
the system is in the ferromagnetic phase with spins oriented along
the total magnetic field \cite{Rademake}. In particular, for $\Omega\gg\Delta,|C_{ij}|$,
all spins align preferentially along the $x$-axis resulting in the
ground state $\ket G=\ket{\rightarrow}_{1}...\ket{\rightarrow}_{N}$,
with $\ket{\rightarrow}=(\ket{\uparrow}-\ket{\downarrow})/\sqrt{(}2)$.


\subsection{Spin-1 model}

\label{sec:spin_one}

The second model is for spin-1 particles, with three internal atomic
states representing the spin-one states $\ket{-1}$, $\ket 0$ and
$\ket 1$ (see section~\ref{sec:imple} for details), described by
the following Hamiltonian  
\begin{eqnarray}
H &= \Delta E\sum_{i}^{N}(1-S_z^{(i)^2})+ \sum_{i<j}^{N}C_{ij}\left(S_{z}^{(i)}S_{z}^{(j)} +(\beta S_{+}^{(i)}S_{-}^{(j)}+h.c.)\right) \\ \nonumber
+&\sum_{i<j}^{N}C_{ij}\left(\gamma S_{+}^{(i)}(S_{z}^{(j)}+S_{z}^{(i)})S_{-}^{(j)}+\delta S_{+}^{(i)}S_{z}^{(i)}S_{z}^{(j)}S_{-}^{(j)}+h.c.\right), \\
 \nonumber
\label{eq:hamil_one}
\end{eqnarray}
where $N$ is the number of spins, $S_{z}^{(i)}=\ket 1_{i}\bra 1_{i}-\ket{-1}_{i}\bra{-1}_{i}$,
$S_{+}^{(i)}=\sqrt{2}(\ket 1\bra 0+\ket 0\bra{-1})$ and $S_{-}^{(i)}=(S_{+}^{(i)})^{\dagger}$
are spin one operators, $\beta$, $\gamma$ and $\delta$ are constants depending on chosen atomic states in the implementation scheme (see section~\ref{sec:imple_one}). The first term in this Hamiltonian corresponds to
an external ``magnetic'' field which acts only on the $0$-th component
of the spin and serves as a controlling parameter during state preparation, while the last two terms represent the DDI including a XXZ-type Hamiltonian (the second term) and higher order spin operators (the last term).  
Obviously {[}c.f. figure~\ref{fig:lattice}{]}, the state $\ket{G'}=\ket 0_{1}...\ket 0_{N}$
minimises the energy of the Hamiltonian for large negative $\Delta E$,
$\Delta E\ll-\left|C_{ij}\right|$. For small $\Delta E$ ($\left|\Delta E\right|\ll\left|C_{ij}\right|$),
the spin exchange term of the Hamiltonian leads to a disordered phase.
Finally, for large positive $\Delta E$, $\Delta E\gg\left|C_{ij}\right|$,
the DDI arranges spins into the stripe phase.


\section{Dynamical preparation of the stripe phase}

\label{sec:result}

We now proceed with the description of the stripe-phase preparation
protocols for both the above models, which is based on the adiabatic
evolution of the ground state~\cite{Vermersch,Pohl,Schachenmayer,Schau=00003D00003D00003D0000DF}.
This is achieved by changing adiabatically the magnetic field from
the values at which the ground state is simple and easily realisable
in experiments, to the values at which the stripe phase is the ground
state. To characterise the state of the system during the protocol,
we introduce the stripe order parameters along the $z$ and $y$ lattice
directions {[}c.f. figure~\ref{fig:lattice}(b){]} 
\begin{equation}
M_{z}=\frac{1}{N}\sum_{i}\langle S_{z}^{(i)}S_{z}^{(i+1_{z})}\rangle,\quad M_{y}=\frac{1}{N}\sum_{i}\langle S_{z}^{(i)}S_{z}^{(i+1_{y})}\rangle,
\end{equation}
where $S_{z}^{(i)}$ are the spin operators of the corresponding model
($S_{z}^{(i)}\rightarrow\sigma_{z}^{(i)}$ for the spin-$1/2$ case),
and $1_{\alpha}$ represents a shift by one site along the $\alpha$-direction.
The system is in the stripe phase when $M_{z}\rightarrow1$ and $M_{y}\rightarrow-1$.

\subsection{Dynamical state preparation of stripe phases for the spin-$1/2$
model}

\label{sec:result_one_half}

For the spin-$1/2$ model, we consider initially (at time $t=t_{0}$)
the system with $\Delta(t_{0})=0$, $\Omega(t_{0})\gg\left|c_{d}\right|$
in its ground state $\ket G$, which can be experimentally prepared
with a high fidelity, and then adiabatically decrease $\Omega$ to
reach the final point $\Delta(t_{f})=0$, $\Omega(t_{f})=0$ at time
$t_{f}$, where the stripe phase is the ground state. Note, however,
that during the parameter ramp, we cross the phase transition point.
Therefore, strictly speaking the criterion of adiabaticity can only
be fulfilled in a finite system where the finite-size effects ensure
the existence of a finite gap even at the transition point. Nevertheless,
keeping in mind that experimental realisations with cold atoms are
always dealing with finite systems, this does not cause any real problems,
but requires the entire time $t_{f}-t_{0}$ of the preparation protocol
be sufficiently large in order to minimise the population of excited
states along the ramp (see below).

We test the protocol with an approach based on a time dependent variational
principle (TDVP) which allows us to simulate the dynamics of the system
with large number of atoms~\cite{Vermersch,Kramer}. This makes possible,
in particular, to estimate the minimal coherence time needed to obtain
the stripe phase experimentally with a good fidelity. The justification
of our approach is based on the comparison with the results of exact
diagonalisation (ED) for small systems.

Within the TDVP approach, the time-evolution of a many-body quantum
state of the spin-$1/2$ model is described via a product state wave-function
\begin{equation}
\ket{\Phi}=\prod_{i=1}^{N}(\alpha_{i}(t)\ket{\uparrow}_{i}+\beta_{i}(t)\ket{\downarrow}_{i}),\label{eq:wavefunc_spin12}
\end{equation}
where $\alpha_{i}(t)$ and $\beta_{i}(t)$ are the time-dependent
coefficients satisfying the normalisation condition $|\alpha_{i}|^{2}+|\beta_{i}|^{2}=1$.
Note that the ground state of the Hamiltonian (\ref{eq:tot_hamiltonian_2level})
with $\Delta=0$ in the classical limit $\Omega\rightarrow0$ (the
stripe phase), is described by this ansatz with the red and blue circles
in figure~\ref{fig:lattice}(b) corresponding to $|\alpha_{i}|^{2}=1,\,\beta_{i}=0$
and $\alpha_{i}=0,\,|\beta_{i}|^{2}=1$, respectively. Our mean field
(MF) ansatz, however, neglects the quantum pair correlation in the
non-classical limit, $\Omega\neq0$.

The time evolution of variational coefficients $\alpha_{i}(t)$ and
$\beta_{i}(t)$ with changing of $\Omega$ and $\Delta$ can be calculated
by the Euler-Lagrange equations~\cite{Vermersch} 
\begin{equation}
\frac{d}{dt}\left(\frac{\partial L}{\partial\dot{\alpha}_{i}^{*}}\right)=\frac{\partial L}{\partial\alpha_{i}^{*}},\label{eq:Euler1}
\end{equation}
\begin{equation}
\frac{d}{dt}\left(\frac{\partial L}{\partial\dot{\beta}_{i}^{*}}\right)=\frac{\partial L}{\partial\beta_{i}^{*}},\label{eq:Euler2}
\end{equation}
where $L$ is the Lagrangian of the system 
\begin{equation}
L=\frac{i}{2}\left(\braket{\Phi}{\dot{\Phi}}-\braket{\dot{\Phi}}{\Phi}\right)-\bra{\Phi}H\ket{\Phi}.\label{eq:lagrangian}
\end{equation}
After substituting the MF ansatz (\ref{eq:wavefunc_spin12}) and the
Hamiltonian (\ref{eq:tot_hamiltonian_2level}) into (\ref{eq:lagrangian}),
the Euler-Lagrange equations (\ref{eq:Euler1}) and (\ref{eq:Euler2})
give 
\begin{equation}
-i\dot{\alpha}_{i}=\left[-\Delta(t)+\sum_{j}C_{ij}(|\beta_{j}|^{2}-|\alpha_{j}|^{2})\right]\alpha_{i}-\Omega(t)\beta_{i},
\label{eq:eq_motion1}
\end{equation}
\begin{equation}
-i\dot{\beta}_{i}=\left[\Delta(t)+\sum_{j}C_{ij}(|\alpha_{j}|^{2}-|\beta_{j}|^{2})\right]\beta_{i}-\Omega(t)\alpha_{i}.
\label{eq:eq_motion2}
\end{equation}
For $C_{ij}=0$, equations~(\ref{eq:eq_motion1}) and (\ref{eq:eq_motion2})
describe the single-spin Rabi oscillations with the frequency $\sqrt{\Delta^{2}+\Omega^{2}}$.
We see that the interactions $C_{ij}$ in the MF description generate
nonlinear corrections to the longitudinal magnetic field $\Delta$
so that the rate of local spin flips depends on the spin configuration
on the neighbouring sites.

\begin{figure}[t]
\centering 
\includegraphics[width=0.8\textwidth]{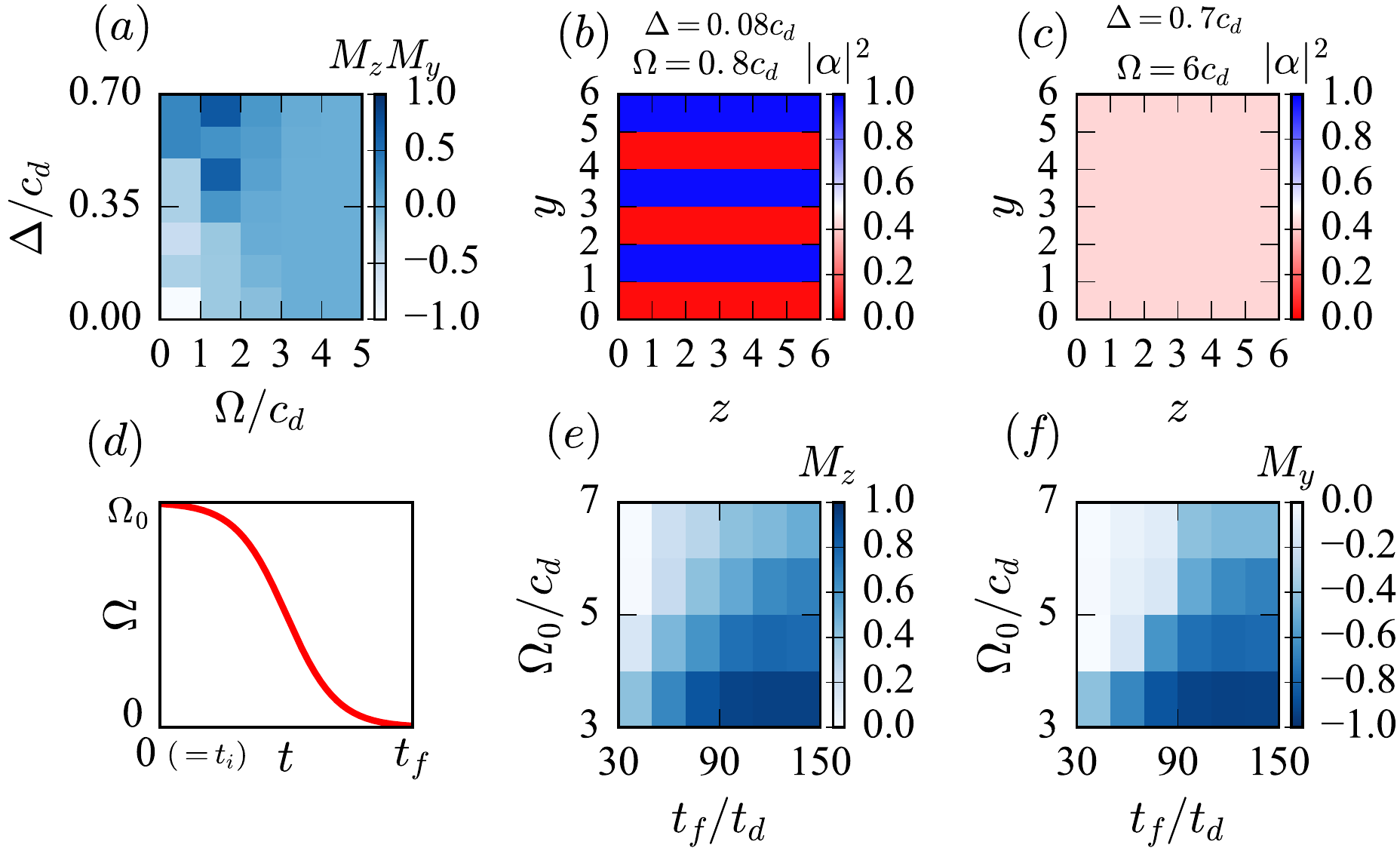}
\caption{The ground state of the spin-$1/2$ model on a $6\times6$ lattice
and the preparation of the stripe phase. (a) The order parameter of
the system $M_{z}M_{y}$ as a function of $\Delta$ and $\Omega$,
indicating the stripe phase for small $\Delta$ and $\Omega$. The
occupation of the state $\protect\ket{\uparrow} $ is shown in panels (b) for
small $\Delta$ and $\Omega$ representing the stripe phase and (c) for large $\Delta$ and $\Omega$ indicating a ferromagnetic phase.
(d) The variation of $\Omega$ with time ($t_{d}=\hbar/c_{d}$) during
the preparation protocol, and the resulting spin-spin correlations
$M_{z}$ and $M_{y}$ at the end of the protocol are presented in
panels (e) and (f), respectively, as a function of the preparation
time $t_{f}$ and the initial value of $\Omega_{0}$. \label{fig:MF_Dyn_2D}}
\end{figure}

The ansatz in equation~(\ref{eq:wavefunc_spin12}) can also be used
to investigate the ground state of the system by considering the parameters
$\alpha_{i}$ and $\beta_{i}$ as independent variational parameters
(subjected to the normalisation constraint only) and minimising the energy of the system. In figure~\ref{fig:MF_Dyn_2D}(a)
we present the results of such calculations for a $6\times6$ lattice
by showing the quantity $M_{z}M_{y}$ as a function of $\Delta$ and
$\Omega$. As expected, the stripe phase ($M_{z}M_{y}\rightarrow-1$)
is the variational ground state of the system when the DDI is larger
than the transverse and longitudinal magnetic fields $\Delta$ and
$\Omega$ {[}c.f. figure~\ref{fig:MF_Dyn_2D}(b){]}. With increasing
magnetic fields, the stripe phase disappears {[}c.f. figure~\ref{fig:MF_Dyn_2D}(c){]},
and for large $\Omega$ and $\Delta$ the ground state is the ferromagnetic
one (for the parameters in figure~\ref{fig:MF_Dyn_2D} (c), the spins
are mostly oriented in the $x$-direction, $S_{x}=N^{-1}\sum_{i}\left\langle \sigma_{x}^{(i)}\right\rangle =-0.96$).

We now analyse the dynamical preparation of the stripe phase along
the trajectory shown in figure~\ref{fig:lattice}(b) with $\Delta=0$
for all times and $\Omega$ decreasing from its initial value $\Omega=\Omega_{0}$
to 0, as represented in figure~\ref{fig:MF_Dyn_2D}(d). For the system
initialised in the state $\ket G=\ket{\rightarrow}_{1}...\ket{\rightarrow}_{N}$,
with $\ket{\rightarrow}=(\ket{\uparrow}-\ket{\downarrow})/\sqrt{2}$,
figures~ \ref{fig:MF_Dyn_2D}(e) and (f) present the final values
of $M_{z}$ and $M_{y}$ as a function of $\Omega_{0}$ and $t_{f}$.
We see that the larger initial value $\Omega_{0}$ the longer
it takes to reach the stripe phase with high fidelity. For $\Omega_{0}=3c_{d}$
the stripe phase is reached already for $t_{f}\sim10^{2}t_{d}$, where
$t_{d}=\hbar/c_{d}$.

\begin{figure}[t]
\centering \includegraphics[width=0.6\textwidth]{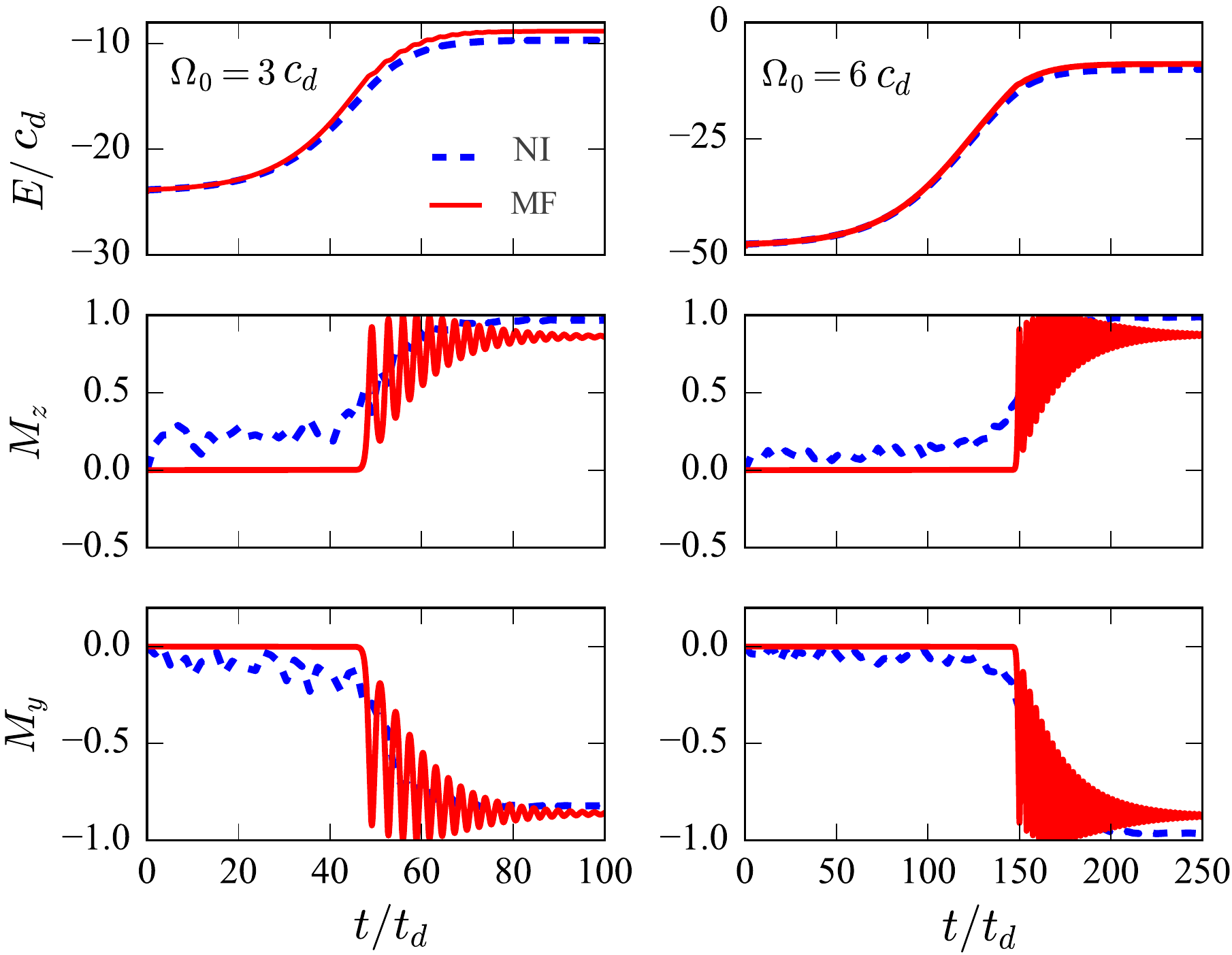}
\caption{Comparison of the results of the dynamical MF approach (solid lines)
with the exact solution of the time-dependent Schr\"odinger equation
(dashed lines) for a $4\times4$ lattice. Time evolution ($t_{d}=\hbar/c_{d}$)
of the energy (top), $M_{z}$ (middle), and $M_{y}$ (bottom) for
different values of $\Omega_{0}$. \label{fig:MF_ED}}
\end{figure}

To validate our MF approach we compare it with the exact numerical
solutions of the time-dependent Schr\"odinger equation for the same
initial state $\ket G=\ket{\rightarrow}_{1}...\ket{\rightarrow}_{N}$
on the $4\times4$ lattice ($N=16$). The comparison is presented
in figure~\ref{fig:MF_ED}, and shows that the two methods are in
a very good agreement when comparing the energy. The order parameters
(spin-spin correlators) $M_{z,y}$ are in agreement within 10\% for
all times except those close to the transition point, where the TDVP
simulations show strong oscillations. They correspond to a large-amplitude
motion of the system over a practically flat potential landscape in
the low-energy manifold of states near the transition. Away from the
transition point, when the system chooses one of the two possible
stripe configurations (see also the discussion below), the oscillations
disappear. In contrast, the numerical solutions of the Schr\"odinger
equation do not show any noticeable oscillations. We attribute this
difference to the fact that the evolution within the ansatz (\ref{eq:wavefunc_spin12})
allows to reach only a very limited (as compared to the true quantum-mechanical
evolution) part of the Hilbert space, making impossible the smearing out
of the total contribution of various oscillating terms in the spin-spin
correlation functions.

\begin{figure}[b]
\centering \includegraphics[width=0.6\textwidth]{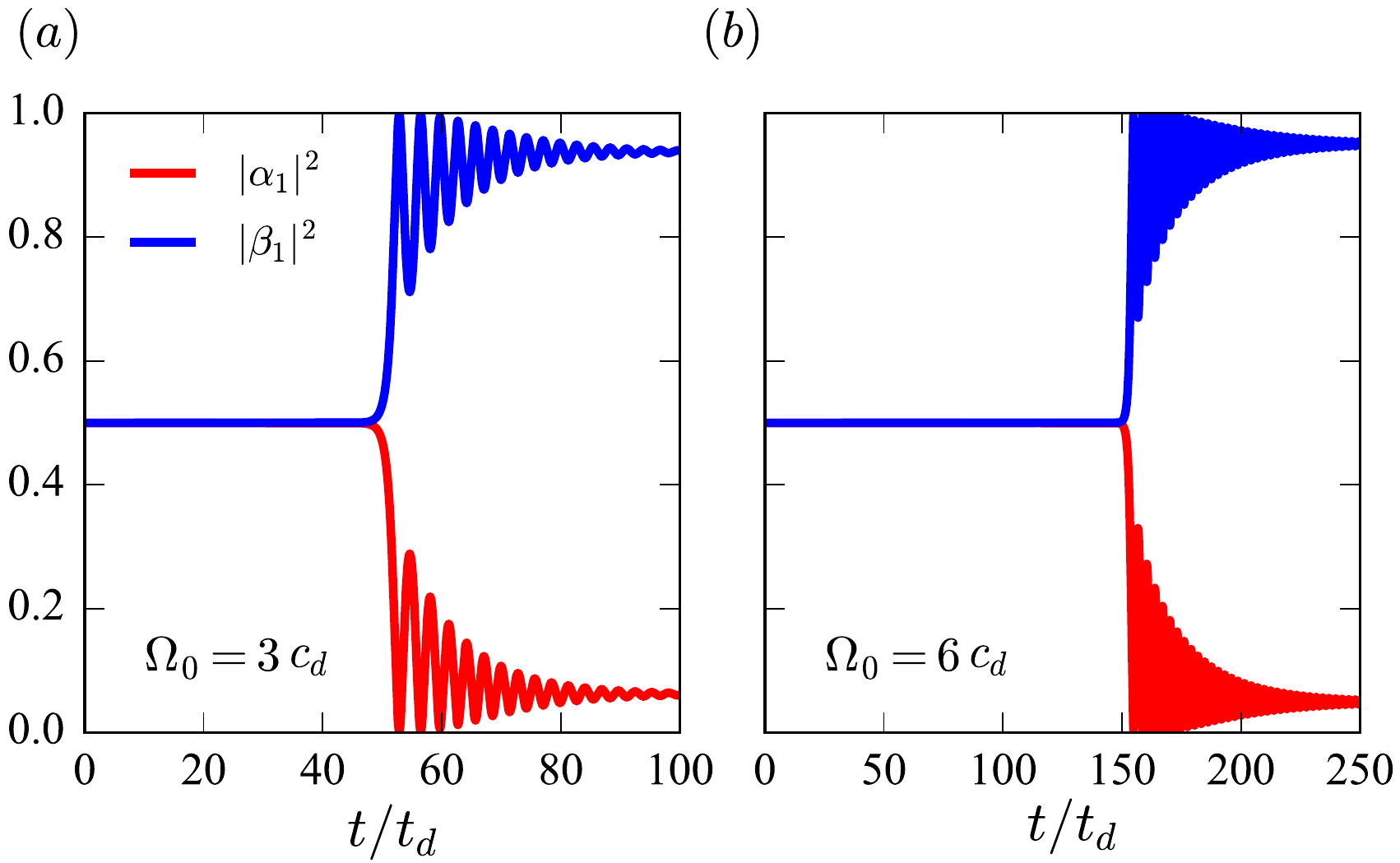} \caption{Time evolution of the occupation of the states $\protect\ket{\uparrow}$
(red line) and $\protect\ket{\downarrow}$ (blue line) in a lattice
site and for different values of $\Omega_{0}$. \label{fig:variational}}
\end{figure}

In figure~\ref{fig:variational}, we have plotted the evolution of
the occupation probabilities of the states $\ket{\uparrow}$ (red
curve) and $\ket{\downarrow}$ (blue curve) to illustrate the dynamics
of the formation of the stripe pattern within the TDVP. It is important
to mention that within the TDVP the system chooses spontaneously
one of the two possible stripe configurations. We believe that this
spontaneous symmetry breaking is a result of finite numerical precision
during the calculation. In contrast, our numerical integration of
the time-dependent Schr\"odinger equation always results in a superposition
of the two possible stripe configurations, showing therefore no sign
of spontaneous symmetry breaking. Such ``robustness'' of the numerical
solutions of the Schr\"odinger equation is due the quantum correlations
(entanglement) over a large part of the Hilbert space, which is built
in the system during the purely quantum-mechanical time evolution.
In real physical systems, however, decoherence processes (due to finite
temperature, noise, etc.) destroy most of these correlations and cause
spontaneous symmetry breaking.

As a final comment we note that, as we have already seen in figure~\ref{fig:MF_Dyn_2D},
the larger $\Omega_{0}$ is, the longer ramps are required to prepare
the stripe phase.

\subsection{Dynamical state preparation of stripe phases for the spin-1 model}

\label{sec:result_one}

\label{sec:result_one}

For the preparation of the stripe phase within the spin-$1$ model
(\ref{eq:hamil_one}), we start with $\Delta E_{0}=\Delta E(t_{i})\ll-\left|C_{ij}\right|$
in the Hamiltonian and initialise the system in the state $\ket{G'}=\ket 0_{1}...\ket 0_{N}$.
We then increase $\Delta E$ adiabatically to a large positive value
$\Delta E_{f}=\Delta E(t_{f})\gg|C_{i,j}|$ {[}c.f. figure~\ref{fig:lattice}(d){]}
to end up in the stripe phase formed on the manifold of spin states
$\ket 1_{i}$ and $\ket{-1}_{i}$ (the spin states $\ket 0_{i}$ are
effectively eliminated by the large positive $\Delta E_{f}$).

Before proceeding to the analysis of the time-evolution of the system
under the ramp we note that the MF variational ansatz used above for
the spin-$1/2$ model, can not be applied here because it is unable
to describe the flip-flop processes of the type $\ket{00}\longleftrightarrow\ket{1,-1}$
involving two spins, see~\ref{app:appendixA}. We thus simulate the
dynamics of the spin-1 model via a time-dependent cluster mean field
(CMF) approach with clusters containing two spins along the $y$-axis
{[}c.f. figure~\ref{fig:lattice}(a){]}. The wave function of the
$\mu$-th cluster is written as 
\begin{equation}
\ket{\psi}_{\mu}=\sum_{l,l^{\prime}}\alpha_{ll^{\prime}}^{(\mu)}\ket l_{1\mu}\ket{l^{\prime}}_{2\mu},\label{eq:cluster_wave}
\end{equation}
where $l,l^{\prime}=-1,\,0,\,1$, the indices $1$ and $2$ label
the particles in the cluster, and $\alpha_{ll^{\prime}}^{(\mu)}$
are the variational parameters subjected to the normalisation constraint
$\sum_{l,l^{\prime}}\left|\alpha_{ll^{\prime}}^{(\mu)}\right|^{2}=1$.
The corresponding ansatz for the wave function of the system is a
product state of all clusters 
\begin{equation}
\ket{\Phi}=\prod_{\mu=1}^{N/2}\ket{\psi}_{\mu}.\label{eq:product_CMF}
\end{equation}
Based on this ansatz and the Hamiltonian (\ref{eq:hamil_one}), we
construct the Lagrangian (\ref{eq:lagrangian}) and the corresponding
Euler-Lagrange equations describing the time evolution of the variational
parameters $\alpha_{ll^{\prime}}^{(\mu)}$, which are rather lengthy
and presented in~\ref{App:AppendixB}.

\begin{figure}[htb]
\centering \includegraphics[width=0.6\textwidth]{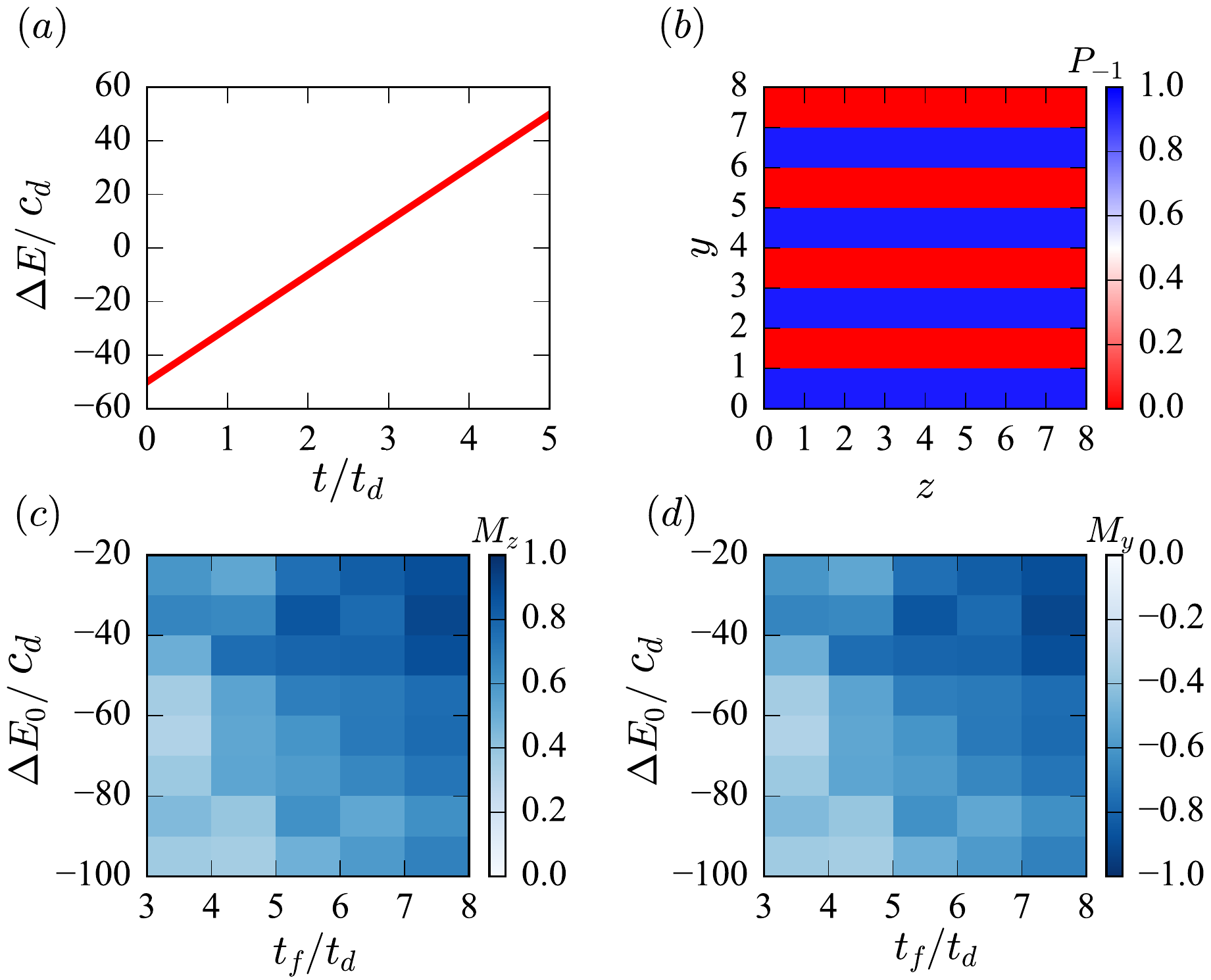} \caption{(a) Time dependence of $\Delta E$ used for the preparation of the
stripe phase. (b) Occupation of state $\ket{-1}$, $P_{-1}$, showing the stripe phase formed on a $8\times8$ lattice at the end of the linear ramp from panel (a). The final values of
the spin-spin correlations (c) $M_{z}$ and (d) $M_{y}$, for the
$8\times8$ lattice as a function of $\Delta E_{0}$ and $t_{f}$
with a fixed value of $\Delta E_{f}=50c_{d}$. \label{fig:CMF_2D}}
\end{figure}

We now apply this scheme to study the dynamical preparation of the
stripe phase with the linear ramp of $\Delta E(t)$. We consider in
the following the parameters of the Hamiltonian~\ref{eq:hamil_one} as $\beta=-2.75$, $\gamma=-1.85$ and $\delta=-1.25$ corresponding
to the case of our implementation with Erbium atoms. In doing the calculations, we also add the term 
\begin{equation}
H_{\mathrm{SB}}=\sum_{\mu}\epsilon(t)\left(\ket 1_{1\mu}\bra 1_{1\mu}-\ket 1_{2\mu}\bra 1_{2\mu}\right)\label{eq:Hsb}
\end{equation}
to the Hamiltonian (\ref{eq:hamil_one}), where $\epsilon(t)$ is
the small energy shift {[}$\epsilon(t_{f})=0${]} between the two
sites (and, therefore, between the states $\ket{-1,1}$ and $\ket{1,-1}$)
in each clusters, to break the up-down symmetry and obtain the stripe
phase {[}c.f.~\ref{App:AppendixB}{]}. Further details and the results
of the calculations on a $8\times8$ lattice are presented in figure~\ref{fig:CMF_2D}.
Panel (a) shows an example of the function $\Delta E(t)$ we used
for the state preparation. In panel (b) we present the occupation
of the state $\ket{-1}$ at the end of evolution starting initially
from the state $\ket{G'}$, which clearly shows the formation of the
stripe phase. For the final value $\Delta E_{f}=50c_{d}$, the spin-spin
correlations $M_{z}$ and $M_{y}$ in the final state are shown in
figures~\ref{fig:CMF_2D}(c) and \ref{fig:CMF_2D}(d), respectively,
as a function of $\Delta E_{0}$ and final time of evolution $t_{f}$.
One clearly sees the effect of the speed of the ramp on the fidelity
of the state preparation.

\begin{figure}[htb]
\centering 
\includegraphics[width=0.6\textwidth]{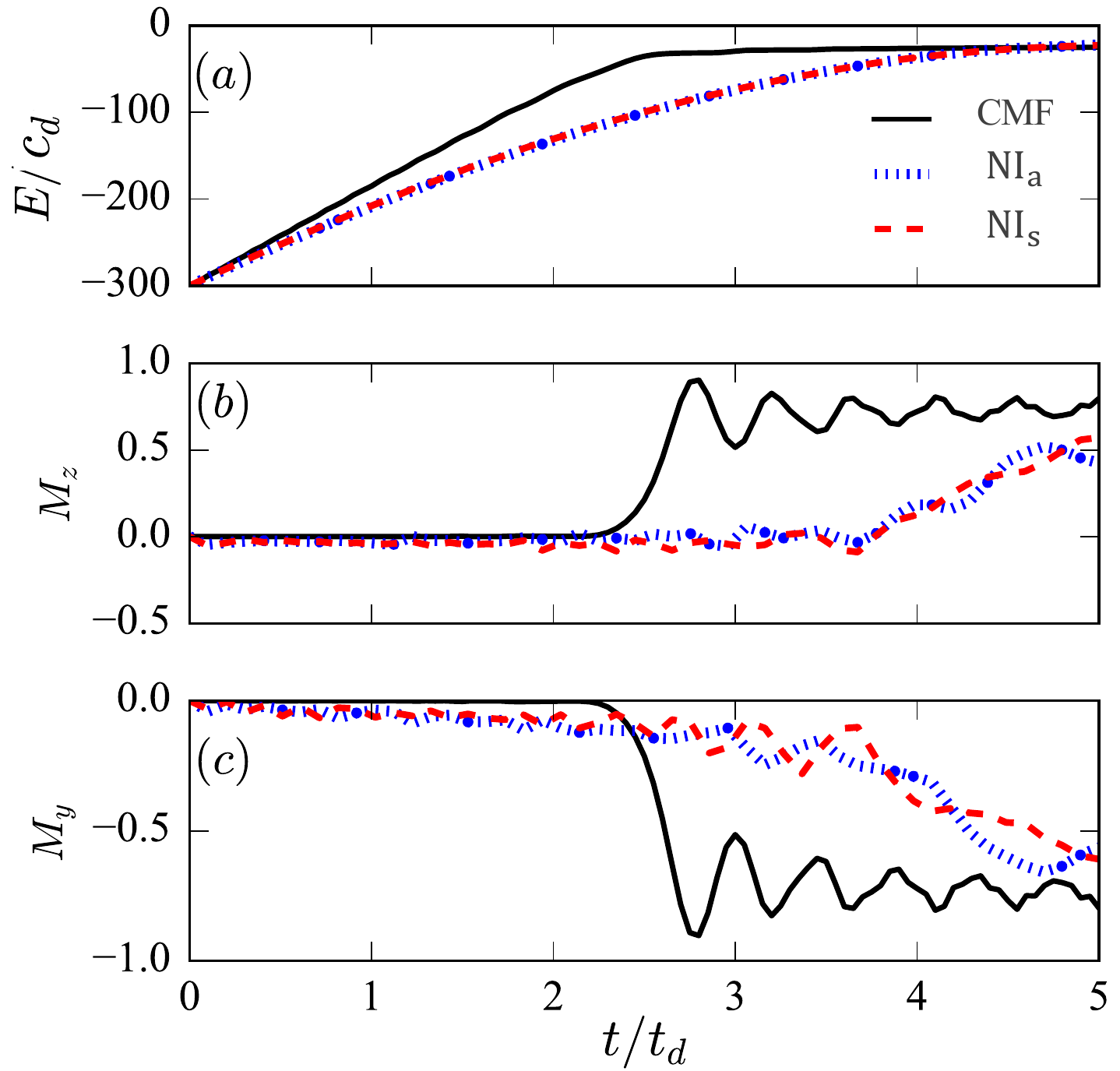} \caption{Time evolution of the (a) energy $E$, (b) $M_{z}$, and (c) $M_{y}$
on a $3\times4$ lattice for the spin-1 model. The solid lines shows
the results of the CMF, the red dashed ($NI_{s}$) and blue dotted
($NI_{a}$) lines are the results of the NI in the presence and absence
of the symmetry-breaking term $H_{\mathrm{SB}}$, respectively. \label{fig:ED_CMF}}
\end{figure}

Finally in figure~\ref{fig:ED_CMF} we compare the results of the
time-dependent CMF (solid lines) with those obtained by direct numerical
integration (NI) of the time-dependent Schr\"odinger equations for the
system on a $3\times4$ lattice. In this figure, the dashed red (dotted
blue) lines are the results of the NI with (without) the symmetry
breaking term $H_{\mathrm{SB}}$ used in the CMF. For all the calculations
we used the linear ramp with $\Delta E$ linearly changing from $\Delta E_{0}=-50c_{d}$
to $\Delta E_{f}=50c_{d}$ between $t_{i}=0$ and $t_{f}=5$.

Concerning the final states of the evolution, the results show good
agreement between the NI and CMF methods: the energies match within
less than one percent, whilst the spin correlations are overestimated
(around twenty percent) by the CMF method. During the evolution, however,
the deviations between the two methods are much more pronounced. This
is not a surprise, keeping in mind the simplicity of the wave function
in the CMF ansatz, equations (\ref{eq:cluster_wave}) and (\ref{eq:product_CMF}),
and have the same origin (very limited part of the Hilbert space used
in the CMF ansatz) as in the case of spin-$1/2$ model discussed above.

\section{Microscopic realisation of spin models with magnetic atoms}

\label{sec:imple} In this section we present an implementation of
aforementioned spin models on the basis of magnetic atoms, with spin
states being encoded in the Zeeman sublevels of the atomic ground
state. We note that, being based on generic properties of the DDIs,
the models can also be realised in other experimental platforms such
as polar molecules~\cite{Yan,Manmana}, Rydberg atoms~\cite{Saffman,Weimer_Nat},
and trapped ions~\cite{Kim,Richerm,Jurcevic}.

\begin{figure}
\centering \includegraphics[width=0.87\textwidth]{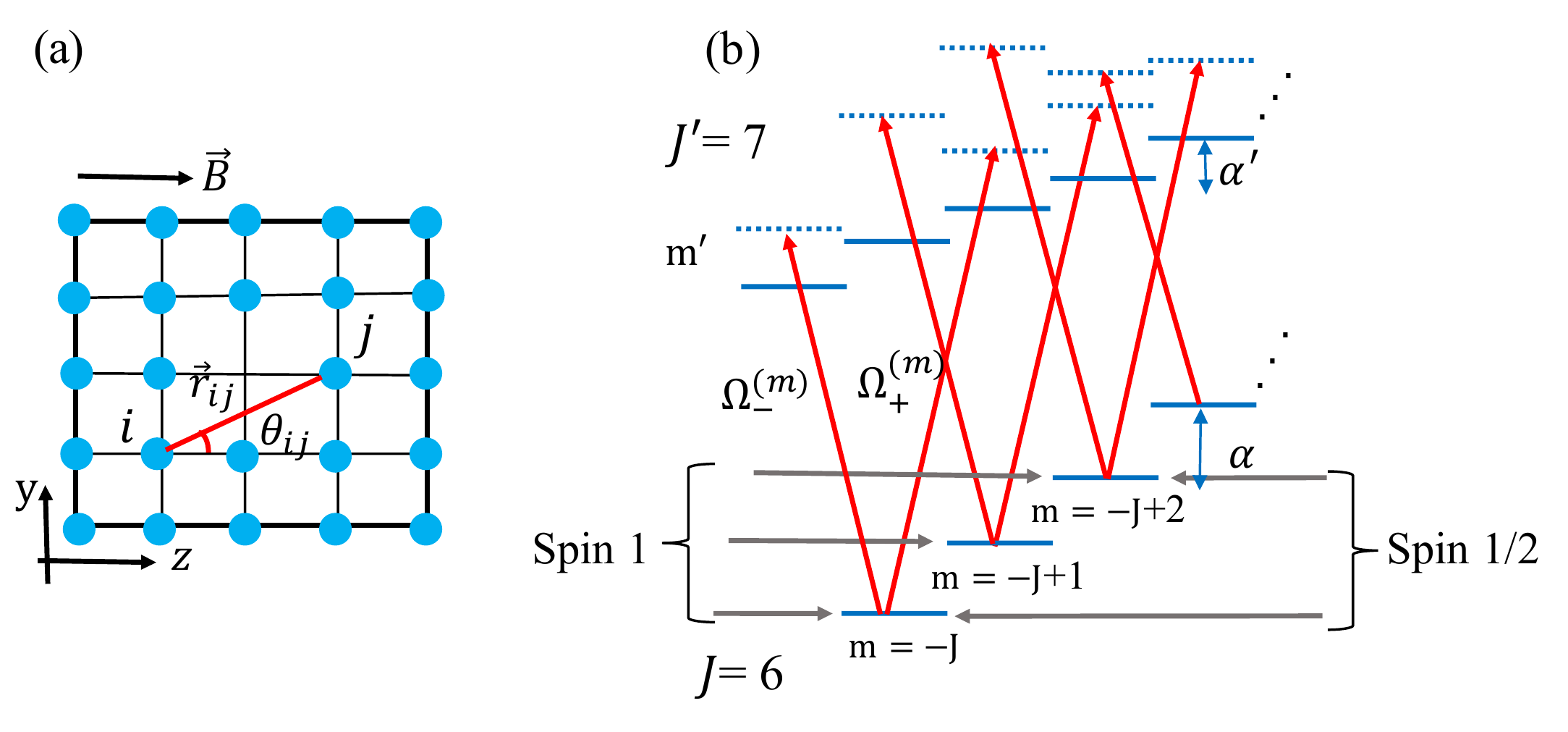} \caption{ (a) Magnetic atoms loaded on a square lattice in the presence of
an in-plane magnetic field. (b) Level structure of a bosonic Erbium atom taken as an example, with the indication of states for realising spin-1 and spin-$1/2$ models. Two right- and left-handed circularly polarised laser beams with $m$-dependent Rabi frequencies $\Omega_{\pm}^{(m)}$
are used to implement state dependent energy shifts. \label{fig:3level_model}}
\end{figure}

The setup we have in mind is represented schematically in figure~\ref{fig:3level_model}
with magnetic atoms placed on a square lattice and prepared in their
electronic ground state manifold. Considering the case of a fine structure
manifold with angular momentum $J$~\footnote{We consider for simplicity the absence of hyperfine interactions.},
our two models will involve a restricted number of levels ($2$ and
$3$), compared to the total number of states $2J+1$ {[}c.f. figure~\ref{fig:3level_model}
(b){]}.

We start with reviewing basic properties of the DDI between magnetic
atoms, then show how to select energetically a particular subset of
Zeeman levels to implement our two spin models. Finally, we discuss
the resulting relevant time scales for the stripe phase preparation.

\subsection{Dipole-dipole interactions between magnetic atoms}

For two magnetic atoms, which are located at sites $i$ and $j$ of
a two-dimensional square lattice, the operator of the DDI reads~\cite{Lahaye2009}
\begin{equation}
V_{ij}=\frac{c_{d}}{r_{ij}^{3}}\left(\boldsymbol{J}^{(i)}.\boldsymbol{J}^{(j)}-3(\boldsymbol{J}^{(i)}.\hat{\boldsymbol{r}}_{ij})(\boldsymbol{J}^{(j)}.\hat{\boldsymbol{r}}_{ij})\right)\label{eq:dd_interaction}
\end{equation}
with $c_{d}=\mu_{0}(g_{J}\mu_{B})^{2}/(4\pi a^{3})$, where $\mu_{0}$
is the magnetic permeability of the vacuum, $\mu_{B}$ is the Bohr
magneton, $g_{J}$ is the Land\'e factor, $J^{i(j)}$ denotes the operator
of the total angular momentum of the particle $i\,(j)$, and $\boldsymbol{r}_{ij}=\boldsymbol{r}_{i}-\boldsymbol{r}_{j}$
is the relative position of two particles ($\hat{\boldsymbol{r}}_{ij}=\boldsymbol{r}_{ij}/r_{ij}$)
in units of the lattice spacing $a$. Assuming the presence of a strong
external magnetic field $\mathbf{B}=B\mathbf{\hat{z}}$ and choosing
its direction as the quantisation axis, the DDI Hamiltonian can be
written as~\cite{Santos2006} 
\begin{equation}
H_{{\rm {int}}}=H_{q=0}+H_{q=1}+H_{q=2},\label{eq:Htot}
\end{equation}
where the first term 
\begin{equation}
H_{q=0}=\sum_{i<j}J_{ij}\left(J_{z}^{(i)}J_{z}^{(j)}-\frac{1}{4}(J_{+}^{(i)}J_{-}^{(j)}+h.c.)\right)
\label{eq:int_spherical_0}
\end{equation}
conserves the total angular momentum of two atoms. Here $J_{ij}=(c_{d}/r_{ij}^{3})(1-3\cos^{2}\theta_{ij})$
with $\theta_{ij}$ being the angle between $\boldsymbol{r}_{ij}$
and the $z-$axis. The second and the last terms of the Hamiltonian
(\ref{eq:Htot}) are 
\begin{equation}
H_{q=1}=-\frac{3}{2}\sum_{i\neq j}\frac{c_{d}}{r_{ij}^{3}}\sin\theta_{ij}\cos\theta_{ij}(J_{+}^{(i)}J_{z}^{(j)}e^{-i\phi_{ij}}+h.c.),
\label{eq:int_spherical_1}
\end{equation}
\begin{equation}
H_{q=2}=-\frac{3}{4}\sum_{i<j}\frac{c_{d}}{r_{ij}^{3}}\sin^{2}\theta_{ij}(J_{+}^{(i)}J_{+}^{(j)}e^{-2i\phi_{ij}}+h.c.),
\label{eq:int_spherical_2}
\end{equation}
where $J_{\pm}=J_{x}\pm iJ_{y}$ and $\phi_{ij}$ is the azimuthal
angle of $\boldsymbol{r}_{ij}$ in the $x-y$ plane. These terms do
not conserve energy and transfer one ($H_{q=1}$) or two ($H_{q=2}$)
units of the internal angular momentum of the atoms to or from their
relative orbital motion. In this work, we only consider the term $H_{q=0}$
assuming the Zeeman splitting induced by the magnetic field to be
much larger than the DDI, which makes the contribution of non-angular
momentum conserving terms $H_{q\neq0}$ negligible. 

\subsection{State-dependent energy shifts}

\label{sec:starkshifts}

To exclude energetically all unwanted states from the dynamics, we
use AC Stark shifts in combination with the linear Zeeman shifts.
To this end we consider two polarised laser beams with polarisation
$\sigma_{\pm}$ and same frequency $\omega$, which couple off-resonantly
the states $\ket m\equiv|J,m_{J}\rangle$ of the ground state manifold
to a manifold of excited states $\ket{m'}\equiv|J',m_{J}'\rangle$,
{[}c.f. figure~\ref{fig:3level_model}{]}. The corresponding Hamiltonian
within the rotating wave approximation and in the frame rotating with
the laser frequency $\omega$ is given by 
\begin{eqnarray}
h & = & \sum_{m^{\prime}=-J^{\prime}}^{J^{\prime}}(\alpha^{\prime}m^{\prime}-\Delta)\ket{m^{\prime}}\bra{m^{\prime}}+\sum_{m=-J}^{J}\alpha m\ket m\bra m\\
 & + & \sum_{m}\Omega_{\pm}^{(m)}(\ket{(m\pm1)'}\bra m+h.c),
\nonumber 
\end{eqnarray}
where different Land\'e factors in the two manifolds give rise to different
Zeeman shifts $\alpha^{(\prime)}=\mu_{B}g_{J^{(\prime)}}B$. We note
that the Rabi frequencies $\Omega_{\pm}^{(m)}=\Omega_{\pm}C_{\pm}^{(m)}$
depend on \textit{m} due to the different Clebsch-Gordan coefficients
$C_{\pm}^{(m)}$ involved in the electric dipole transitions. Finally,
$\Delta=\omega-\omega_{0}$ is the detuning where $\omega_{0}$ denotes
the resonant frequency between the ground and the first excited states
in the absence of the magnetic field. Taking the laser coupling as
a perturbation ($|\Delta-\alpha^{\prime}(m\pm1)+\alpha m|\gg\Omega_{\pm}^{(m)}$),
the effective Hamiltonian for the ground state manifold in the second
order perturbation theory can be written as 
\begin{equation}
h_J=\sum_{m=-J}^{J}(\alpha m+\epsilon_{m})\ket m\bra m,\label{eq:hamil_spin1}
\end{equation}
where $\epsilon_{m}=\frac{\Omega_{+}^{(m)^{2}}}{\Delta-\alpha^{\prime}(m+1)+\alpha m}+\frac{\Omega_{-}^{(m)^{2}}}{\Delta-\alpha^{\prime}(m-1)+\alpha m}$.
(Note that we neglected the two photon Raman transitions $|m\rangle\to|m+2\rangle$
assuming $\alpha, \Delta\gg\Omega_{\pm}^{(m)}$.) The AC stark-shifts
$\epsilon_{m}$ depend non-linearly on $m$, which will allow energy-conserving
dynamics only on a restricted set of two (three) Zeeman states in
the ground state manifold for the spin $1/2$ (spin $1$) model. We also
note that an alternative way to realise the nonlinear energy
shifts is by using the quadratic Zeeman effect.

\subsection{Implementation of the spin-1 model}

\label{sec:imple_one}

\begin{figure}
\centering \includegraphics[width=0.6\textwidth]{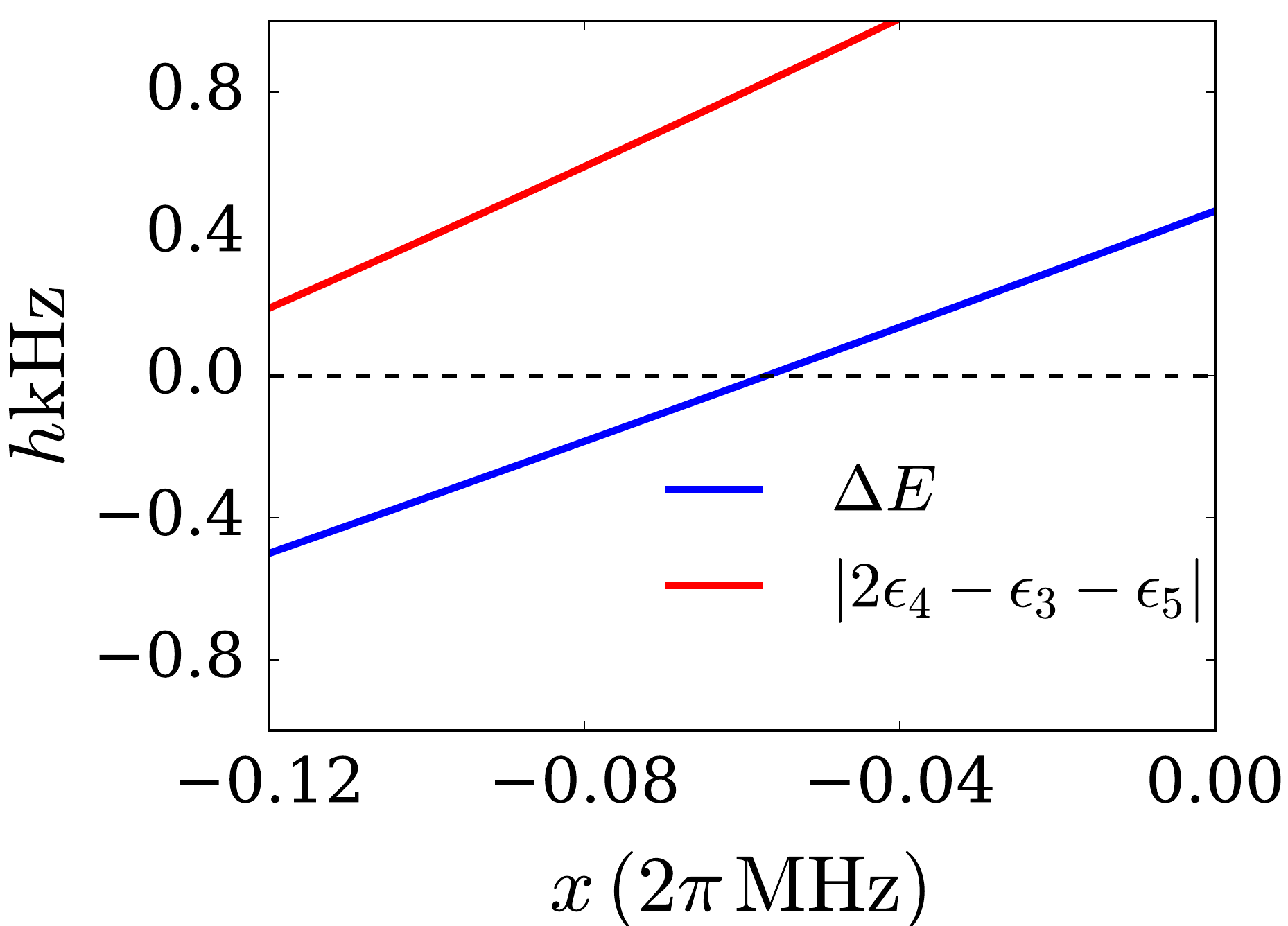} \caption{(blue line) The control parameter in the spin-1 model, $\Delta E$, for Erbium atoms, which changes from a negative value to a positive one to prepare the stripe phase at the end. (red line) the energy shift for the transition $\ket{m=-4,m=-4}\longleftrightarrow\ket{m=-5,m=-3}$, which is forbidden due to being much larger than $c_d=2\pi\times0.9~\hbar\mathrm{Hz}$. Note that here $\Omega_{\pm}=2\pi\times0.5~\mathrm{MHz}\pm x$, $\Delta=0.8\alpha$ and the applied magnetic field is $B=5~\mathrm{G}$. \label{fig:energy_shifts}}
\end{figure}

Following the discussion above, the effective Hamiltonian governing
the dynamics on the ground state manifold has the form $H=H_{q=0}+h_{J}$.
To implement the spin-1 model from section~\ref{sec:spin_one}, we
consider, for example, the three adjacent magnetic levels $\ket{m=m_{0}-1,m_{0},m_{0}+1}$
with $m_{0}=-J+1$ {[}c.f. figure~\ref{fig:3level_model}{]} (another
option would be $m_{0}=J-1$), and impose the condition $|2\epsilon_{m_{0}+1}-\epsilon_{m_{0}}-\epsilon_{m_{0}+2}|\gg c_{d}$, which suppresses the process $\ket{m_{0}+1,m_{0}+1}\to\ket{m_{0},m_{0}+2}$
(due to $H_{q=0}$) energetically. This allows us to project the Hamiltonian
$H$ onto the submanifold with $m=m_{0}-1,m_{0},m_{0}+1$ defining
the spin-one states $\ket k$ ($k=-1,0,1$) as $\ket k=\ket{m=m_{0}+k}e^{-i(\alpha(m_{0}+k)+\epsilon_{m_{0}+k}-\Delta E/2\delta_{k,0})t}$ with $\Delta E=2\epsilon_{m_{0}}-\epsilon_{m_{0}-1}-\epsilon_{m_{0}+1}$. One then gets the parameters $\beta=-J_1^2/8$, $\gamma=-J_1(J_1-J_2)/8$ and $\delta=-(J_1-J_2)^2/8$ with $J_1=\sqrt{J(J+1)-m_0(m_0+1)}$ and $J_2=\sqrt{J(J+1)-m_0(m_0-1)}$.
Note that the energy shift $\Delta E$ can be varied in time by using
the time-dependent Rabi frequencies $\Omega_{\pm}$ and the detuning
$\Delta$. As an example, we consider ground state bosonic Erbium atoms with $J=6$~\footnote{The hyperfine-structure is absent in this case.} ($g_{J}=1.1638$)~\cite{Conway},
which are trapped in a square lattice with lattice spacing $a=266$
nm~\cite{Baier}. This results in the dipole-dipole coupling $c_{d}=2\pi\times0.9~\hbar\mathrm{Hz}$. 
In order to satisfy the requirements above for preparing the stripe phase for the case of Erbium atoms with $m_0=-5$, one can apply  two laser beams with the Rabi frequencies $\Omega_{\pm}=2\pi\times0.5~\mathrm{MHz}\pm x$ and the detuning $\Delta=0.8\alpha$, which couple the ground state manifold to the manifold with $J'=7$ ($g_{J'}=1.070$). In the presence of a magnetic field $B=5~\mathrm{G}$ ($\alpha=51.186~\hbar\mathrm{MHz}$ and $\alpha^\prime=47.06~\hbar\mathrm{MHz}$), we get figure~\ref{fig:energy_shifts}, which shows all the requirements are fulfilled.

Finally, we emphasise that the spin-$1$ model only involves magnetisation
conserving terms and is thus robust (in first order in the Zeeman
shifts) against time-dependent fluctuations of the magnetic field.
This is not the case for the spin 1/2 model discussed below.

\subsection{Implementation of the spin-$1/2$ model}

\label{sec:imple_half}

The spin-$1/2$ model can be implemented by choosing two states $\ket{m=-J}$,
$\ket{m=-J+2}$ in the ground state manifold as spin states, $\ket{-J}=\ket{\uparrow}$
and $\ket{-J+2}=\ket{\downarrow}$ {[}c.f. figure~\ref{fig:3level_model}{]}.
To exclude the other Zeeman states from the dynamics, we require the
conditions $|\epsilon_{-J}+\epsilon_{-J+2}-2\epsilon_{-J+1}|\gg c_{d}$
and $|\epsilon_{-J+3}+\epsilon_{-J+1}-2\epsilon_{-J+2}|\gg c_{d}$.
The resulting projection of the Hamiltonian $H_{q=0}$ onto the submanifold
$\ket{-J+2},\ket{-J}$ leads to the first term in~(\ref{eq:tot_hamiltonian_2level}).
To obtain the other terms we can use an additional Raman coupling
of the two states $\ket{-J}$ and $\ket{-J+2}$. In the rotating wave-approximation,
this coupling takes the form of the two magnetic fields $\Omega$
and $\Delta$ which appear in (\ref{eq:tot_hamiltonian_2level})~\footnote{Note that we can neglect the transition between other ground state
levels $|m\rangle$ assuming that the Land\'e factors between ground
($J$) and excited manifolds allow to make state-selective Raman transition
between $\ket{-J+2}$ and $\ket{-J}$}. For the purpose of the adiabatic state preparation, these fields
can be controlled by changing the Rabi-frequency and the detuning
of the additional Raman coupling (see, for example,~\cite{Schau=00003D00003D00003D0000DF}).


\subsection{Time scales}

\label{scalling} With the above implementation schemes in mind, we
can now estimate the time scales required for our state-preparation
protocol under realistic experimental conditions. Coming back to the previously discussed example, i.e. Erbium atoms, one gets $2\pi\times3.66~\hbar\mathrm{Hz}$ for the exchange
interaction. This gives $t_f<1\mathrm{s}$ for the time required
to adiabatically prepare the stripe phase in the spin-1 case ($t_d\sim0.177\mathrm{s}$),
which is well compatible with experimental time scales. (For the spin
1/2 case, the required time is order of magnitude larger, which does
not look realistic.) We also note that the Raman couplings involved
in the implementation of the models give rise to decoherence with
the rate $\Gamma\sim(\frac{\Omega}{\Delta})^{2}\Gamma_{r}$, where
$\Gamma_{r}$ is the line width of the excited states involved. This
rate should be much smaller than the exchange interaction, which requires
$\Delta\gg\Omega,\Gamma_{r}$~\cite{Frisch2,Marek,Frisch1}.

\section{Conclusion}

\label{sec:conclusion}

We have shown how systems of magnetic atoms in optical lattices can
be used to realise stripe phases in current experiments. On the methodological
side, we have employed time-dependent variational ans\"atze which, as
also reported in~\cite{Vermersch}, seem particularly well suited
to address adiabatic state preparation for magnetically ordered states.
Most importantly, they enable the investigation of systems sizes comparable
with current cold atom experiments, that is well beyond what can
be reached with exact methods. We remark that our many-body results
can find immediate applications to other physical settings, such as
polar molecules and Rydberg atoms in optical lattices~\cite{Lahaye2009},
superconducting qubits~\cite{Dalmonte2015,Viehmann}, where both
Heisenberg-type and Ising Hamiltonians with dipolar interactions can
be realised.

Our results show how the combination of long coherence times and flexibility
in initialising and manipulating magnetic atoms represent an ideal
tool to explore quantum magnetism, despite the interaction strengths
being weaker (in absolute value) with respect to polar molecules and
Rydberg gases. Moreover, the tunability of the magnetic field orientation/strength
and the properties of the DDIs also allow in principle to realise
other interesting quantum phases, such as integer or fractional Chern
insulators~\cite{Yao_prl,Yao,Syzranov_nat,Peter,Syzranov}, for which
it would be intriguing to formulate a proper time-dependent variational
principle to correctly reproduce the exact dynamics.

\section*{Acknowlegments}

We acknowledge inspiring and fruitful discussions with P. Zoller. We also thank S. Baier, J. H. Becher, L. Chomaz, F. Ferlaino, M. J. Mark, and G. Natale for helpful discussions on several aspects of the implementation schemes. The exact diagonalisation simulations were performed using
the QuTiP toolbox~\cite{Johansson}. Research in Innsbruck is supported
by the European Research Council (ERC) Synergy Grant UQUAM, EU H2020
FET Proactive projects RySQ, and by the Austrian Science Fund through
SFB FOQUS (FWF Project No. F4016-N23).


\section*{References}

\bibliographystyle{unsrtnat}


\appendix

\section{Equations of motion of the spin-1 model within time-dependent mean
field}

\label{app:appendixA} Here we demonstrate that the time dependent
MF approach for the spin-1 model cannot describe the dynamics of the
system. Having three available states $\ket{-1}$, $\ket 0$ and $\ket 1$
for each atom, one can start from the product state as

\begin{equation}
\ket{\Phi}=\prod_{i=1}^{N}\sum_{l}\alpha_{l,i}\ket l_{i},\label{eq:wave_MF}
\end{equation}
where $l=-1,0,1$ and the coefficients $\alpha_{l,i}$ are the variational
parameter of the state $\ket l$ of the spin $i$. With this ansatz,
the Euler-Lagrange equations for the Lagrangian (\ref{eq:lagrangian})
corresponding to the Hamiltonian (\ref{eq:hamil_one}) read

\begin{eqnarray}
-i\dot{\alpha}_{-1}^{i}&=&\alpha_{-1}^{i}\sum_{j}C_{ij}(|\alpha_{1}^{j}|^{2}-|\alpha_{-1}^{j}|^{2})+\frac{\alpha_{0}^{i}}{4}\sum_{j}C_{ij}\left(J_{2}^{2}\alpha_{0}^{j*}\alpha_{-1}^{j}+J_{1}J_{2}\alpha_{1}^{j*}\alpha_{0}^{j}\right)\nonumber \\
-i\dot{\alpha}_{0}^{i}&=&-\alpha_{0}^{i}\Delta E+\frac{\alpha_{1}^{i}}{4}\sum_{j}C_{ij}\left(J_{1}^{2}\alpha_{1}^{j*}\alpha_{0}^{j}+J_{1}J_{2}\alpha_{0}^{j*}\alpha_{-1}^{j}\right)\nonumber \\
&+&\frac{\alpha_{-1}^{i}}{4}\sum_{j}C_{ij}\left(J_{2}^{2}\alpha_{-1}^{j*}\alpha_{0}^{j}+J_{1}J_{2}\alpha_{0}^{j*}\alpha_{1}^{j}\right)\nonumber \\
-i\dot{\alpha}_{1}^{i}&=&\alpha_{1}^{i}\sum_{j}C_{ij}(|\alpha_{-1}^{j}|^{2}-|\alpha_{1}^{j}|^{2})
\nonumber \\
&+&\frac{\alpha_{0}^{i}}{4}\sum_{j}C_{ij}\left(J_{1}^{2}\alpha_{0}^{j*}\alpha_{1}^{j}
J_{1}J_{2}\alpha_{-1}^{j*}\alpha_{0}^{j}\right).
\end{eqnarray}

Note that in the Hamiltonian (\ref{eq:hamil_one}), we have used $\beta=-J_1^2/8$, $\gamma=-J_1(J_1-J_2)/8$ and $\delta=-(J_1-J_2)^2/8$ with $J_1=\sqrt{22}$ and $J_2=\sqrt{12}$.
It is easy to see that the state $\ket{G'}$ (the initial state in
our stripe phase preparation) with $\alpha_{\pm1}^{i}(t_{i})=0$ and
$\alpha_{0}^{i}(t_{i})=1$ is the eigenstate of the above system of
equations: $\ket{G'(t)}=\exp[-iN\Delta E(t-t_{i})]\ket{G'}$. On the
other hand, the state $\ket{G'}$ is not the eigenstate of the Hamiltonian
(\ref{eq:hamil_one}), because of the flip-flop processes of the type
$\ket{00}\leftrightarrow\ket{1,-1}$ resulting from the term $S_{+}^{(i)}S_{-}^{(j)}+S_{-}^{(i)}S_{+}^{(j)}$
in (\ref{eq:hamil_one}). Being not able to incorporate these processes,
the MF approach fails in describing the evolution of the system.

\section{Dynamics of the spin-1 model within time-dependent cluster mean field}

\label{App:AppendixB}

We here present details about the dynamical CMF approach for the spin-1
model within the ansatz (\ref{eq:product_CMF}). The corresponding
Euler-Lagrange equations can be written in the following matrix form

\begin{equation}
\frac{d}{dt}\alpha_{\mu}=C\alpha_{\mu},
\end{equation}
with $\alpha_{\mu}$ being the $9$-component vector made of the variational
parameters for the $\mu$-th cluster {[}see (\ref{eq:cluster_wave}){]},
$\alpha_{\mu}=(\alpha_{1,1},\alpha_{1,0},\alpha_{1,-1},\alpha_{0,1},\alpha_{0,0},\alpha_{0,-1},\alpha_{-1,1},\alpha_{-1,0},\alpha_{-1,-1})_{\mu}^{\dagger}$,
and
 
\begin{equation}
C=\left[\begin{array}{ccccccccc}
C_{0} & J_{1}A^{*} & 0 & J_{1}B^{*} & 0 & 0 & 0 & 0 & 0\\
J_{1}A & C_{1} & J_{2}A^{*} & \frac{J_{1}^{2}}{4} & J_{1}B^{*} & 0 & 0 & 0 & 0\\
0 & J_{2}A & C_{2} & 0 & \frac{J_{1}J_{2}}{4} & J_{1}B^{*} & 0 & 0 & 0\\
J_{1}B & \frac{J_{1}^{2}}{4} & 0 & C_{3} & J_{1}A^{*} & 0 & J_{2}B^{*} & 0 & 0\\
0 & J_{1}B & \frac{J_{1}J_{2}}{4} & J_{1}A & C_{4} & J_{2}A^{*} & \frac{J_{1}J_{2}}{4} & J_{2}B^{*} & 0\\
0 & 0 & J_{1}B & 0 & J_{2}A & C_{5} & 0 & \frac{J_{2}^{2}}{4} & J_{2}B^{*}\\
0 & 0 & 0 & J_{2}B & \frac{J_{1}J_{2}}{4} & 0 & C_{6} & J_{1}A^{*} & 0\\
0 & 0 & 0 & 0 & J_{2}B & \frac{J_{2}^{2}}{4} & J_{1}A & C_{7} & J_{2}A^{*}\\
0 & 0 & 0 & 0 & 0 & J_{2}B & 0 & J_{2}A & C_{8}
\end{array}\right],
\end{equation}
where 
\begin{eqnarray}
A & =\sum_{\nu}\left(\frac{\gamma_{3\nu}}{2}(C_{\mu\nu}^{21}+C_{\mu\nu}^{12})+C_{\mu\nu}^{22}\gamma_{4\nu}\right)\nonumber \\
B & =\sum_{\nu}\left(\frac{\gamma_{4\nu}}{2}(C_{\mu\nu}^{21}+C_{\mu\nu}^{12})+C_{\mu\nu}^{11}\gamma_{3\nu}\right)\nonumber \\
C_{0} & =\sum_{\nu}\left(-\frac{\gamma_{1\nu}+\gamma_{2\nu}}{2}(C_{\mu\nu}^{21}+C_{\mu\nu}^{12})-C_{\mu\nu}^{11}\gamma_{1\nu}-C_{\mu\nu}^{22}\gamma_{2\nu}\right)-1\nonumber \\
C_{1} & =\sum_{\nu}\left(-C_{\mu\nu}^{11}\gamma_{1\nu}-\frac{\gamma_{2\nu}}{2}(C_{\mu\nu}^{21}+C_{\mu\nu}^{12})\right)-\Delta E+\epsilon\nonumber \\
C_{2} & =\sum_{\nu}\left(-\frac{(\gamma_{2\nu}-\gamma_{1\nu})}{2}(C_{\mu\nu}^{21}+C_{\mu\nu}^{12})-C_{\mu\nu}^{11}\gamma_{1\nu}+C_{\mu\nu}^{22}\gamma_{2\nu}\right)+1+\epsilon\nonumber \\
C_{3} & =\sum_{\nu}\left(-C_{\mu\nu}^{22}\gamma_{2\nu}-\frac{\gamma_{1\nu}}{2}(C_{\mu\nu}^{21}+C_{\mu\nu}^{12})\right)-\Delta E-\epsilon\nonumber \\
C_{4} & =-2\Delta E\nonumber \\
C_{5} & =\sum_{\nu}\left(C_{\mu\nu}^{22}\gamma_{2\nu}+\frac{\gamma_{1\nu}}{2}(C_{\mu\nu}^{21}+C_{\mu\nu}^{12})\right)-\Delta E\nonumber \\
C_{6} & =\sum_{\nu}\left(\frac{(\gamma_{2\nu}-\gamma_{1\nu})}{2}(C_{\mu\nu}^{21}+C_{\mu\nu}^{12})+C_{\mu\nu}^{11}\gamma_{1\nu}-C_{\mu\nu}^{22}\gamma_{2\nu}\right)+1-\epsilon\nonumber \\
C_{7} & =\sum_{\nu}\left(C_{\mu\nu}^{11}\gamma_{1\nu}+\frac{\gamma_{2\nu}}{2}(C_{\mu\nu}^{21}+C_{\mu\nu}^{12})\right)-\Delta E\nonumber \\
C_{8} & =\sum_{\nu}\left(\frac{(\gamma_{2\nu}+\gamma_{1\nu})}{2}(C_{\mu\nu}^{21}+C_{\mu\nu}^{12})+C_{\mu\nu}^{11}\gamma_{1\nu}+C_{\mu\nu}^{22}\gamma_{2\nu}\right)-1,
\end{eqnarray}
the coefficients $C_{\mu\nu}^{l,l'}=(1-3\cos^{2}\theta_{\mu,\nu}^{l,l^{\prime}})r_{\mu\nu}^{-3}$
correspond to the dipole-dipole coupling between the $l$-th particle
of the $\mu$-th cluster and the $l^{\prime}$-th particle of the
$\nu$-th cluster. We define 

\begin{eqnarray}
\gamma_{1\nu} & =\sum_{l=-1}^{1}{(|\alpha_{1,l}^{\nu}|^{2}-|\alpha_{-1,l}^{\nu}|^{2})}\nonumber \\
\gamma_{2\nu} & =\sum_{l=-1}^{1}{(|\alpha_{l,1}^{\nu}|^{2}-|\alpha_{l,-1}^{\nu}|^{2})}\nonumber \\
\gamma_{3\nu} & =\frac{1}{4}\sum_{l=-1}^{1}{(J_{2}\alpha_{-1,l}^{\nu}\alpha_{0,l}^{\nu^{*}}+J_{1}\alpha_{0,l}^{\nu}\alpha_{1,l}^{\nu^{*}})}\nonumber \\
\gamma_{4\nu} & =\frac{1}{4}\sum_{l=-1}^{1}{(J_{2}\alpha_{l,-1}^{\nu}\alpha_{l,0}^{\nu^{*}}+J_{1}\alpha_{l,0}^{\nu}\alpha_{l,1}^{\nu^{*}})}.
\end{eqnarray}

Note that in section~\ref{sec:result_one} we have introduced (time-dependent,
see main text) energy difference $\epsilon$ between the states $\ket{1,-1}$
and $\ket{-1,1}$ in each cluster to break the symmetry between them,
which prevents the dynamical formation of the stripe phase. The inclusion
of $\epsilon$, for which we assume a decaying time dependence, breaks
the translational symmetry and results in the formation of the stripe
phase. 
\end{document}